\documentclass[12pt]{iopart}
\usepackage{epsfig} 
\usepackage{amssymb}
\begin{document}

\title{Status report of the baseline collimation system of CLIC. Part II}

\author{J.~Resta-L\'opez$^1$, D.~Angal-Kalinin$^2$, B.~Dalena$^3$, J.~L.~Fern\'andez-Hernando$^2$, F.~Jackson$^2$, D.~Schulte$^3$, A.~Seryi$^1$ and R.~Tom\'as$^3$}

\address{$^1$JAI, University of Oxford, UK}
\address{$^2$STFC, Daresbury, UK}
\address{$^3$CERN, Geneva, Switzerland}

\ead{j.restalopez@physics.ox.ac.uk}

\begin{abstract}
Important efforts have recently been dedicated to the characterisation and improvement of the design of the post-linac collimation system of the Compact Linear Collider (CLIC). This system consists of two sections: one dedicated to the collimation of off-energy particles and another one for betatron collimation. The energy collimation system is further conceived as protection system against damage by errant beams. In this respect, special attention is paid to the optimisation of the energy collimator design. The material and the physical parameters of the energy collimators are selected to withstand the impact of an entire bunch train. Concerning the betatron collimation section, different aspects of the design have been optimised: the transverse collimation depths have been recalculated in order to reduce the collimator wakefield effects while maintaining a good efficiency in cleaning the undesired beam halo; the geometric design of the spoilers has been reviewed to minimise wakefields; in addition, the optics design has been optimised to improve the collimation efficiency. This report presents the current status of the the post-linac collimation system of CLIC. Part II is mainly dedicated to the study of the betatron collimation system and collimator wakefield effects.     
\end{abstract}

\maketitle
\section{Introduction}
\label{intro}

The post-linac collimation systems of the future linear colliders will play an essential role in reducing the detector background at the interaction point (IP), and protecting the machine by minimising  the activation and damage of sensitive accelerator components. 

\begin{figure}[htb]
\begin{center}
\includegraphics*[width=14cm]{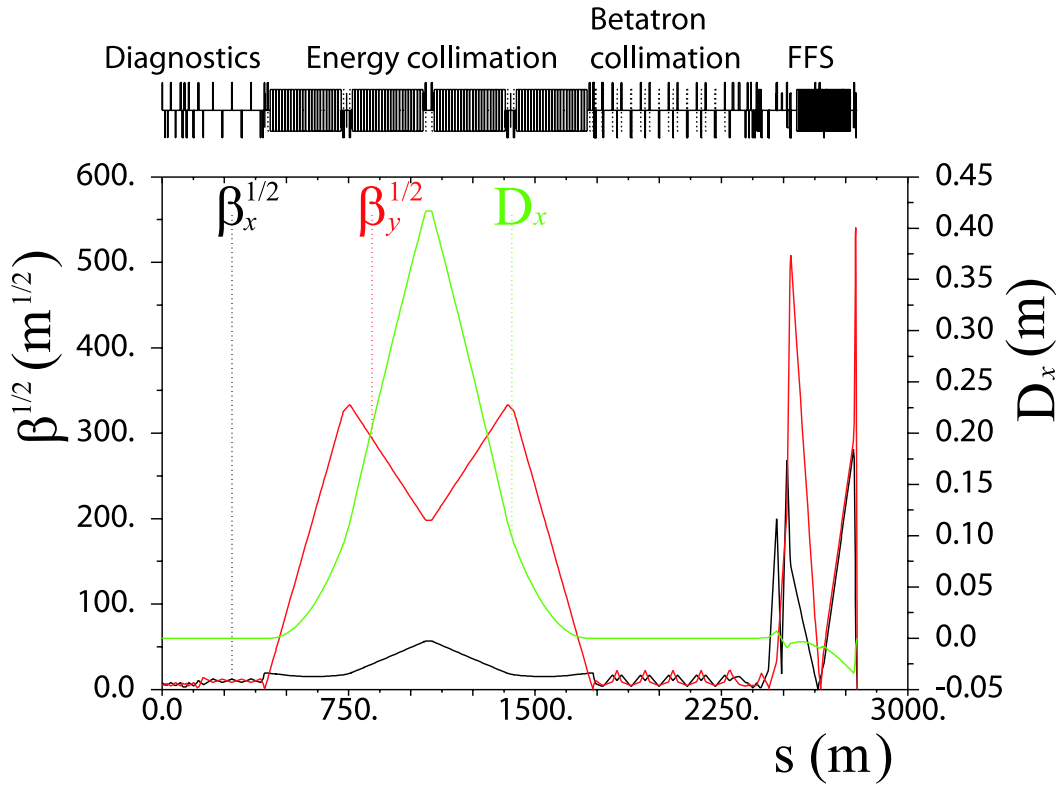}
\caption{Optical functions of the CLIC beam delivery system.}
\label{latticecoll}
\end{center}
\end{figure}

The CLIC Beam Delivery System (BDS), downstream of the main linac, consists of a 370 m long diagnostics section, an almost 2000 m long collimation system, and a 460 m long Final Focus System (FFS) \cite{Rogelio, Tecker}. Figure~\ref{latticecoll} shows the betatron and dispersion functions along the CLIC BDS. Some relevant CLIC design parameters are shown in Table~\ref{CLICparametros} for the options at 500 GeV and 3 TeV centre-of-mass (CM) energy.

\begin{table}[!htb]
\caption{CLIC parameters at 0.5~TeV and 3~TeV CM energy.}
\label{CLICparametros}
\begin{center}
\begin{tabular}{lcc}
\hline
Parameter & {\bf CLIC 0.5~TeV} & {\bf CLIC 3~TeV} \\
\hline \hline
Design luminosity ($10^{34}$ cm$^{-2}$s$^{-1}$) & 2.3 & 5.9 \\
Linac repetition rate (Hz) & 50 & 50  \\
Particles/bunch at IP ($\times 10^{9}$) & 6.8 & 3.72 \\
Bunches/pulse & 354 & 312 \\
Bunch length ($\mu$m) & 72 & 44 \\
Bunch separation (ns) & 0.5 & 0.5 \\
Bunch train length (ns) & 177 & 156 \\
Emittances $\gamma \epsilon_{x}$/$\gamma \epsilon_{y}$ (nm rad) & 2400/25 & 660/20 \\
Transverse beam sizes at IP $\sigma^{*}_x$/$\sigma^{*}_y$ (nm) & 202/2.3 & 45/0.9 \\
BDS length (km) & 1.73 & 2.79 \\
\hline
\end{tabular}
\end{center}
\end{table}

\begin{figure}[htb]
\begin{center}
\includegraphics*[width=14cm]{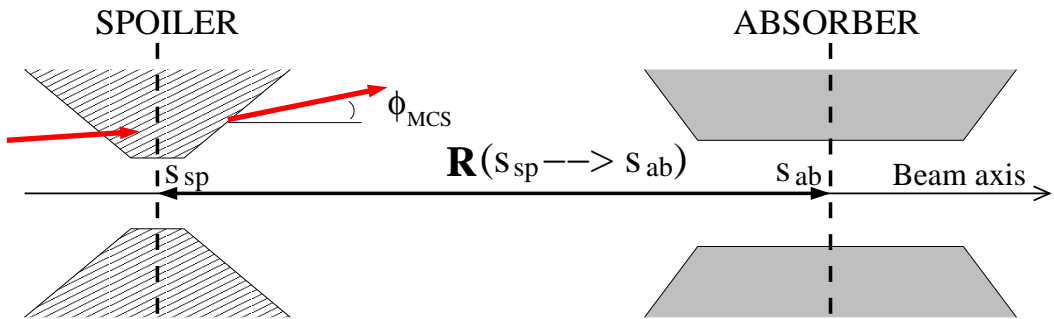}
\caption{Basic spoiler-absorber scheme.}
\label{spoilerabsorberscheme}
\end{center}
\end{figure}

In the CLIC BDS there are two collimation sections:

\begin{itemize}
\item The first post-linac collimation section is dedicated to energy collimation. The energy collimation depth is determined by failure modes in the linac \cite{DanielandFZ}. A spoiler-absorber scheme (Fig.~\ref{spoilerabsorberscheme}), located in a region with non-zero horizontal dispersion, is used for intercepting miss-steered or errant beams with energy deviation larger than $1.3\%$ of the nominal beam energy.    

\item Downstream of the energy collimation section, a dispersion-free section, containing eight spoilers and eight absorbers, is dedicated to the cleaning of the transverse halo of the beam, thereby reducing the experimental background at the IP. 
\end{itemize}

\noindent The spoilers are thin devices ($\lesssim 1$ radiation length) which scrape the beam halo and, if accidentally struck by the full power beam, will increase the volume of the phase space occupied by the incident beam via multiple Coulomb scattering. In this way, the transverse density of the scattered beam is reduced for passive protection of the downstream absorber. The absorbers are usually thick blocks of material (of about 20 radiation length) designed to provide efficient halo absorption or complete removal of potentially dangerous beams.    

The optics of the CLIC collimation system was originally designed by rescaling  of the optics of the collimation system of the previous Next Linear Collider (NLC) project at 1 TeV centre-of-mass energy \cite{PT2, Assmann1} to the 3~TeV CLIC requirements. In the present CLIC baseline optics the length of the energy collimation section has been scaled by a factor 5 and the bending angles by a factor $1/12$ with respect to the 1 TeV NLC design \cite{Assmann2}. On the other hand, the optics of the CLIC betatron collimation section was not modified with respect to the original design of the NLC. 

It is worth mentioning that, unlike the International Linear collider (ILC) \cite{ILC}, where the betatron collimation section is followed by the energy collimators, in CLIC the energy collimation section is upstream of the betatron one. The main reason of choosing this lattice structure is because miss-phased or unstable off-energy drive beams are likely failure modes in CLIC, and they are expected to be much more frequent than large betatron oscillations with small emittance beams. Therefore, the energy collimation system is conceived as the first post-linac line of defence for passive protection against off-energy beams in the CLIC BDS.

Recently many aspects of the CLIC collimation system design have been reviewed and optimised towards a consistent and robust system for the Conceptual Design Report of CLIC (CLIC CDR), to be completed during 2011. In this report we describe the current status of the CLIC collimation system at 3~TeV CM energy. Here we mainly focus on the description of the collimation layout and the optimisation of the necessary parameters of the baseline design to improve the collimation performance, only taking into account the primary beam halo. The aim is to define basic specifications of the design. Studies including secondary particle production and muon collimation are described elsewhere \cite{Agapov, Deacon}. Part II is mainly dedicated to the study of the betatron collimation system and collimator wakefield effects. Moreover, the current status of the CLIC collimation system at 500 GeV CM energy is also described.           

\section{Betatron collimation}
\label{betatronsection}

\begin{figure}[htb]
\begin{center}
\includegraphics*[width=14cm]{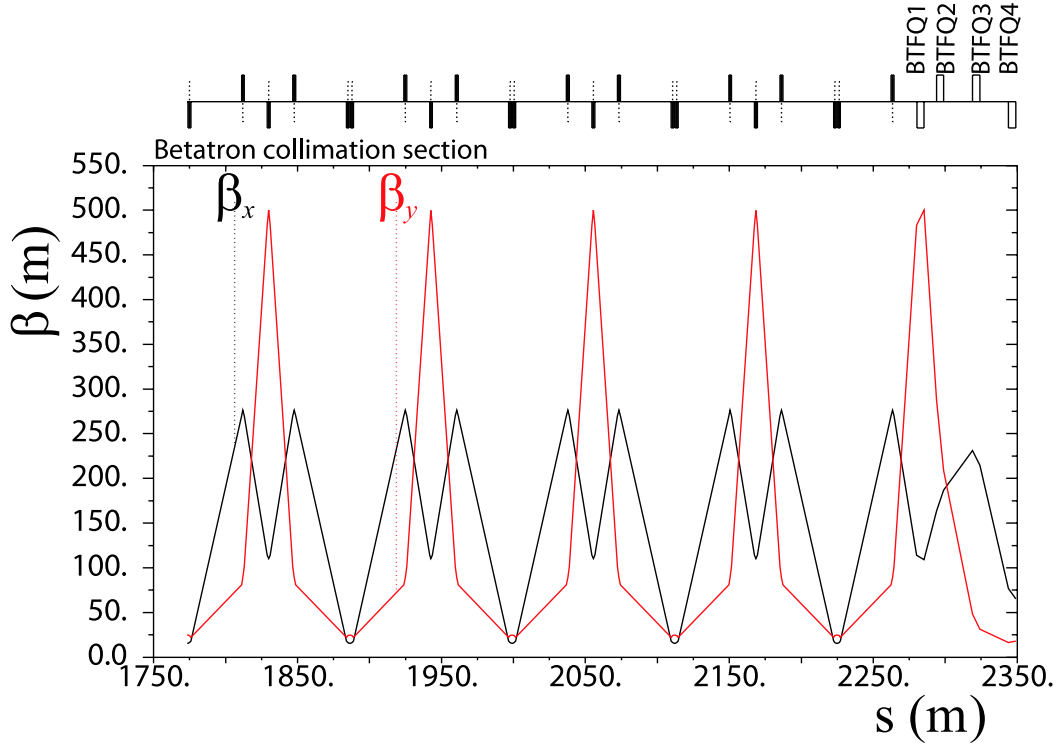}
\caption{Optical functions of the CLIC betatron collimation section.}
\label{latticebetatroncoll}
\end{center}
\end{figure}

The main function of the betatron collimation section is the removal of any particle from the transverse halo of the primary beam, i.e.~beam particles with large betatron amplitudes, which can cause unacceptable experimental background levels in the interaction region. In addition, the collimation system design must limit the regeneration of halo due to optical or collimator wakefield effects. The optics of the betatron collimation section is shown in Fig.~\ref{latticebetatroncoll}. The values of the betatron functions and transverse beam size at each betatron collimator (spoiler and absorber) position are indicated in Table~\ref{tablecoll5}. 

\begin{table}[!htb]
%\begin{sidewaystable}
\caption{Optics and beam parameters at collimator position: longitudinal position, horizontal and vertical $\beta$-functions, horizontal dispersion, horizontal and vertical rms beam sizes. YSP$\#$ denotes vertical spoiler, XSP$\#$ horizontal spoiler, YAB$\#$ vertical absorber and XAB$\#$ horizontal absorber.}
\label{tablecoll5}
\begin{center}
\begin{tabular}{|l|c|c|c|c|c|c|}
\hline 
Name & s [m] & $\beta_{x}$ [m] & $\beta_{y}$ [m] & $D_x$ [m] & $\sigma_x$ [$\mu$m] & $\sigma_y$ [$\mu$m] \\
\hline \hline
YSP1 & 1830.872 &  114.054 &   483.252 &   0. & 5.064 & 1.814 \\ 
XSP1 & 1846.694 &  270.003 &   101.347 &   0. & 7.792 & 0.831 \\
XAB1 & 1923.893 &  270.102 &   80.905  &   0. & 7.793 & 0.742  \\
YAB1 & 1941.715 &  114.054 &   483.185 &   0. & 5.064 & 1.814 \\
YSP2 & 1943.715 &  114.054 &   483.189 &   0. & 5.064 & 1.814 \\
XSP2 & 1959.536 &  270.002 &   101.361 &   0. & 7.791 & 0.831  \\
XAB2 & 2036.736 &  270.105 &    80.944 &   0. & 7.793 & 0.743  \\ 
YAB2 & 2054.558 &  114.054 &   483.255 &   0. & 5.064 & 1.814 \\
YSP3 & 2056.558 &  114.054 &   483.253 &   0. & 5.064 & 1.814 \\
XSP3 & 2072.379 &  270.003 &   101.347 &   0. & 7.791 & 0.831  \\
XAB3 & 2149.579 &  270.102 &    80.905 &   0. & 7.793 & 0.742  \\
YAB3 & 2167.401 &  114.054 &   483.185 &   0. & 5.064 & 1.814 \\
YSP4 & 2169.401 &  114.054 &   483.189 &   0. & 5.064 & 1.814 \\
XSP4 & 2185.222 &  270.002 &   101.361 &   0. & 7.791 & 0.831 \\
XAB4 & 2262.422 &  270.105 &    80.944 &   0. & 7.793 & 0.743  \\
YAB4 & 2280.243 &  114.055 &   483.255 &   0. & 5.064 & 1.814 \\
\hline 
\end{tabular}
\end{center}
%\end{sidewaystable}
\end{table}

In order to provide an acceptable cleaning efficiency of the transverse beam halo the betatron collimation depths are determined from the following conditions: 
\begin{itemize}

\item Minimisation of the synchrotron radiation photons in the first final quadrupole magnet (QF1) that can hit the second final quadrupole (QD0).
%\item The synchrotron radiation photons emitted in the first final quadrupole magnet (QF1) should not hit the second final quadrupole (QD0). 

\item Minimisation of the beam particles that can hit either QF1 or QD0.
%\item No beam particles should hit either QF1 or QD0. 

\item Neither synchrotron radiation photons nor electrons (positrons) of the beam are permitted to impact the detector or its mask.
%\item The collimation apertures should be enough to provide an acceptable cleaning efficiency of the undesired beam halo. 

\end{itemize}

\begin{figure}[htb]
\begin{center}
\includegraphics*[width=14cm]{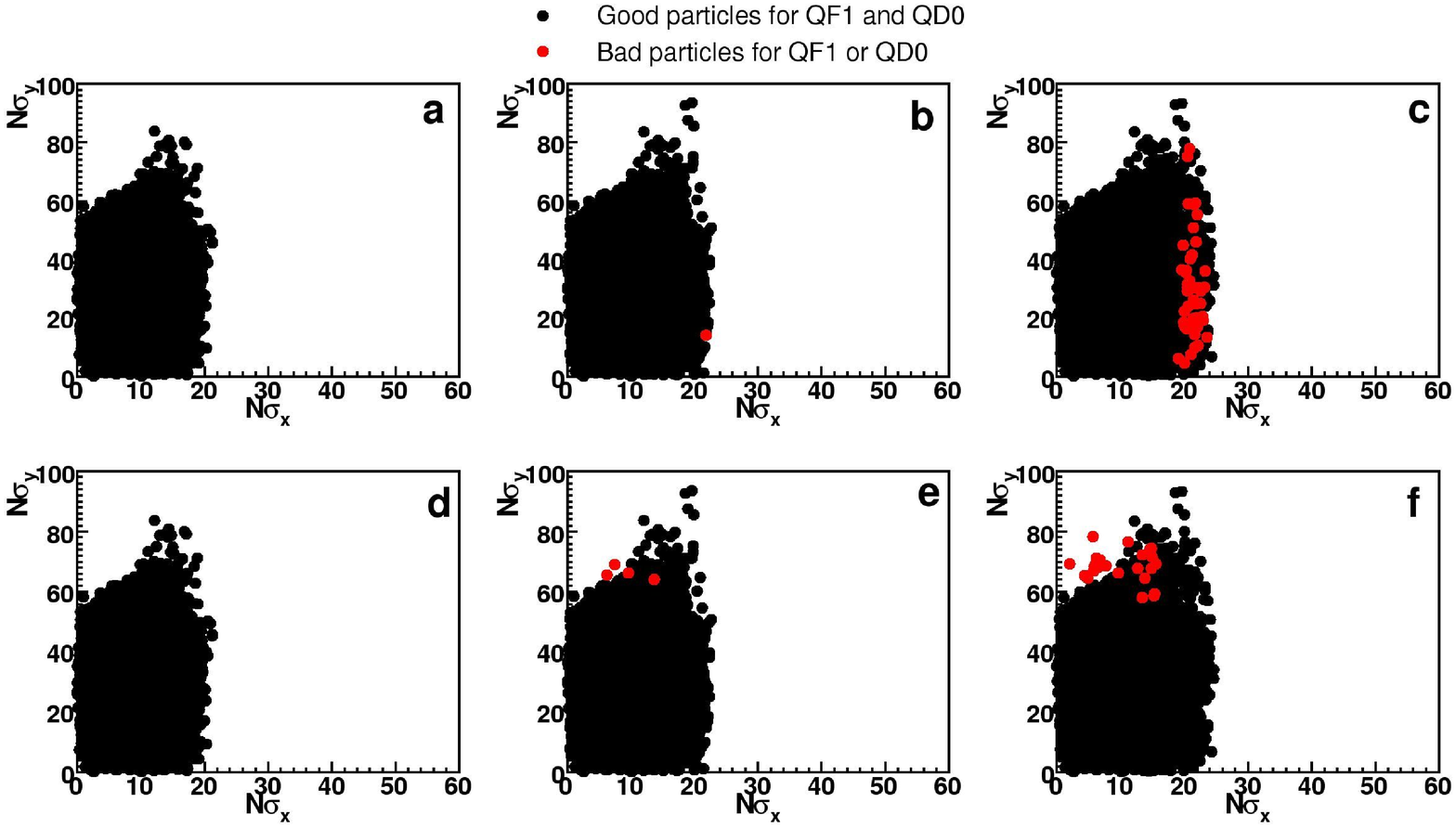}
\caption{(Colour) Transverse beam distribution at the BDS entrance: non-dangerous macroparticles for the final doublet magnets are in black and potentially dangerous macroparticles are in red, according to different collimator apertures. The axes show the position of the particles in number of sigma in the $x$--$x'$ and $y$--$y'$ planes. In the following the corresponding horizontal and vertical collimator apertures (half gaps $a_{x,y}$) are given: a) $a_x=0.11$~mm (13.7~$\sigma_x$) and $a_y=0.08$~mm (44~$\sigma_y$), b) $a_x=0.12$~mm (15~$\sigma_x$) and $a_y=0.08$~mm, c) $a_x=0.13$~mm (16.2~$\sigma_x$) and $a_y=0.08$~mm, d) $a_x=0.08$~mm (10~$\sigma_x$) and $a_y=0.09$~mm (49.5~$\sigma_y$), e) $a_x=0.08$~mm and $a_y=0.10$~mm (50~$\sigma_y$), f) $a_x=0.08$~mm and $a_y=0.11$~mm (60.5~$\sigma_y$).}
\label{scanapert}
\end{center}
\end{figure}

\noindent Macroparticles with high transverse amplitude have been tracked along the CLIC BDS using the code PLACET \cite{placetoctave}, taking into account the emission of synchrotron radiation and all the non-linear elements of the system. The particle positions and angles have been checked at the entrance, in the middle and at the exit of QF1 and QD0. Figure~\ref{scanapert} shows the potentially dangerous particles (in red) according to the above conditions for different collimation apertures. The dangerous particles ("bad particles" in Fig.~\ref{scanapert}), i.e. particles which can generate unacceptable background at the IP, are efficiently removed for collimator aperture $< 15~\sigma_x$ in the horizontal plane and $< 55~\sigma_y$ in the vertical plane. Therefore, we have defined $15~\sigma_x$ and $55~\sigma_y$ as the transverse collimation depths. 

\begin{figure}[htb]
\begin{center}
\includegraphics*[width=14cm]{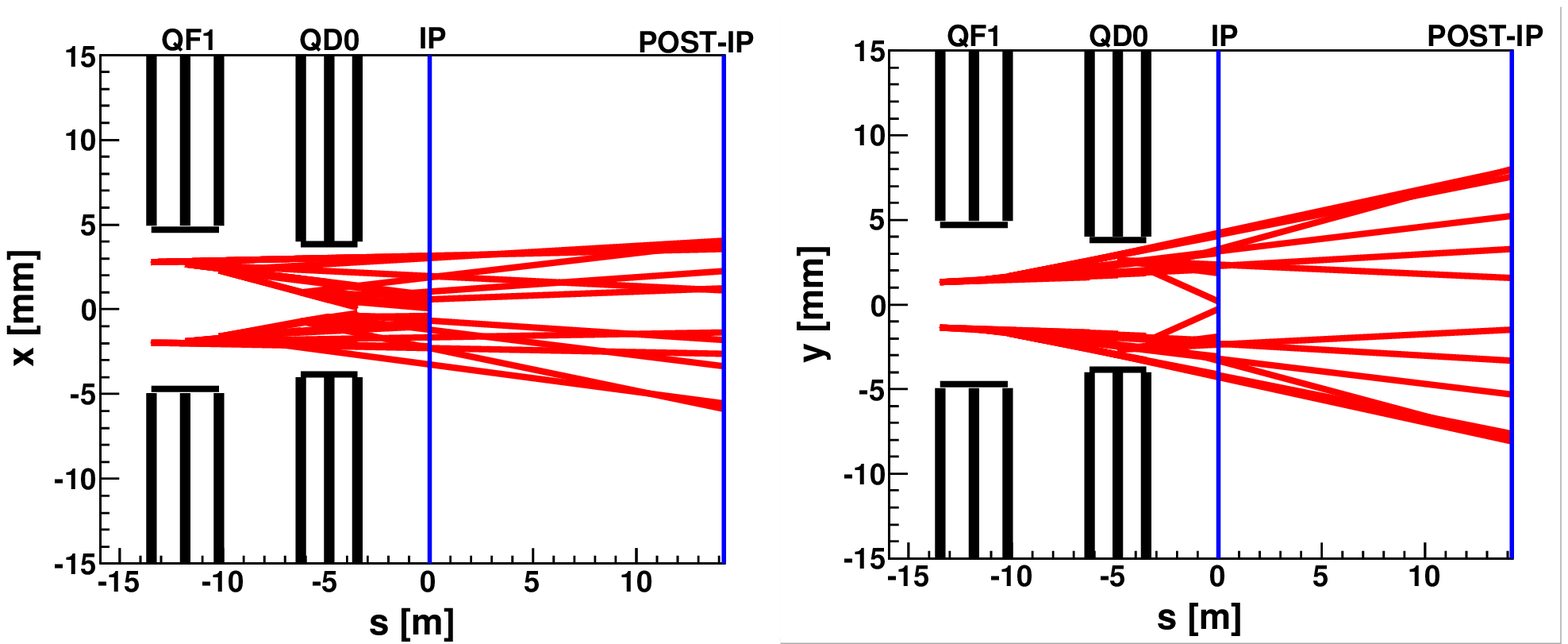}
\caption{Synchrotron radiation fans in the CLIC interaction region emitted by particles with transverse amplitudes $15~\sigma_x$ and $55~\sigma_y$ (the betatron collimation envelope) in the final doublet magnets QF1 and QD0.}
\label{photonfans}
\end{center}
\end{figure}

Figure~\ref{photonfans} shows the residual synchrotron radiation fans from the final quadrupoles QF1 and QD0 to the IP for an envelope covering 15 standard deviations in $x$ and 55 in $y$. At the IP the photon cone is inside a cylinder with radius of 5~mm, which is within the beam pipe radius\footnote{For the CLIC ILD (4 Tesla solenoid) detector configuration \cite{ILD} the inner beam pipe radius at the IP is 29.4 mm, and for the CLIC SiD (5 Tesla solenoid) detector configuration \cite{SiD} the radius is 24.5~mm \cite{AndreSailer}.}. Therefore, in principle, they are not an issue of concern from the detector point of view. 

\begin{figure}[htb]
\begin{center}
\includegraphics*[width=14cm]{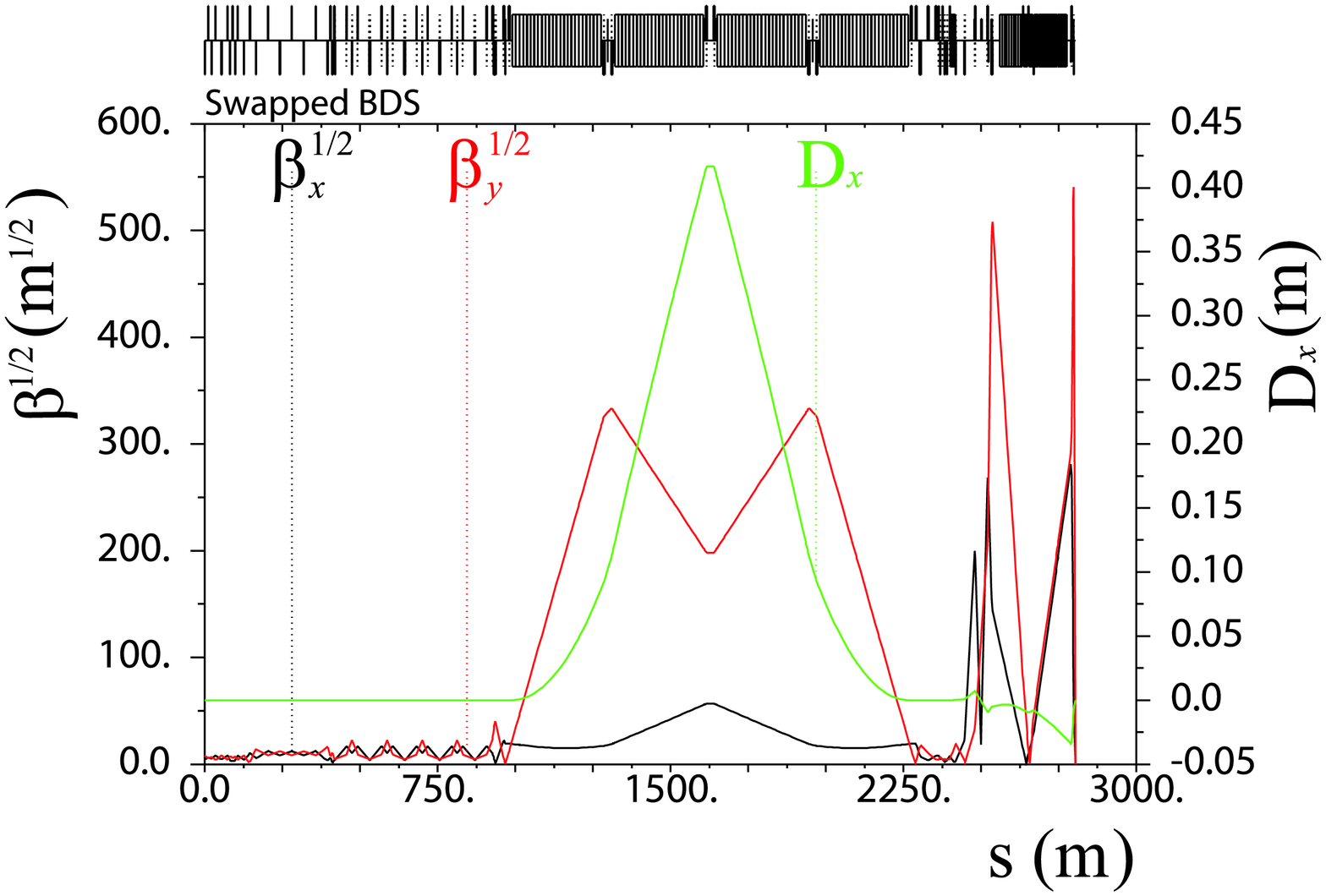}
\caption{Optical functions of the CLIC beam delivery system with swapped lattice, i.e. the betatron collimation system upstream of the energy collimation section.}
\label{swappedlattice}
\end{center}
\end{figure}

It should be considered whether swapping the betatron and energy collimation sections (see Fig.~\ref{swappedlattice}) may lead to further improvement on the betatron cleaning efficiency. This issue has recently been investigated by means of sophisticated tracking simulations, taking into account the halo generation by beam-gas scattering (Mott scattering) and inelastic scattering (Bremsstrahlung) in both linac and BDS, and the production of secondaries \cite{Deacon}. These simulations indicate that the effect of swapping the betatron and energy collimation sections results only in modest $40\%$ reduction in the muon flux reaching the detector. We have decided to maintain the original order of location of the collimation sections in the CLIC BDS. In this way, errant beams coming from the linac would first hit the energy collimators before arriving to the betatron collimation part. In this sense, the energy collimators would protect the betatron collimators of possible damaging.  

%Beam tracking studies, using the code PLACET-Octave, have shown that the effect of swapping the order of the collimation sections has a modest improvement of about $10\%$ in the reduction of the primary halo. More sophisticated tracking simulations , taking into account the halo generation by  

\subsection{Spoiler design and absorber protection}

\begin{figure}[htb]
\begin{center}
\includegraphics*[width=10cm]{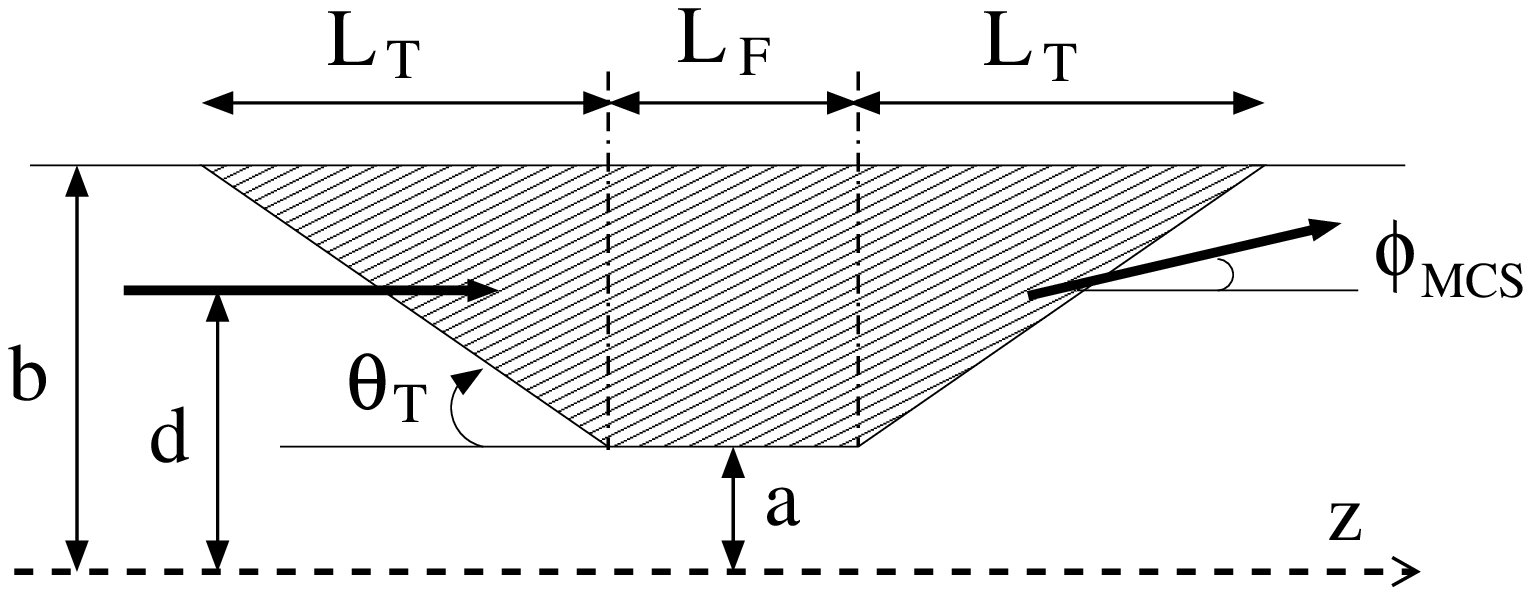}
\caption{Spoiler and absorber jaw longitudinal view.}
\label{spoilergeom}
\end{center}
\end{figure}

The betatron spoilers must scrape the transverse beam halo at the required collimation depths. They must further provide enough beam angular divergence by MCS to decrease the transverse density of an incident beam, thus reducing the damage probability of the downstream absorber. By using similar arguments as in Section~2.1.1 (Part~I), for the protection of the CLIC betatron absorbers, which are made of Ti alloy coated by a thin Cu layer, the rms radial beam size $\sigma_r(s_{ab})=\sqrt{\sigma_x (s_{ab}) \sigma_y (s_{ab})}$ must be larger than about $600~\mu$m at the absorber position \cite{PT2, PT1}. This condition determines the necessary minimum length of the betatron spoiler. The spoilers and absorbers are assumed to have the longitudinal geometry of Fig.~\ref{spoilergeom}.

Considering the linear transport between a betatron spoiler and its corresponding downstream absorber, the expected value of the square of the transverse displacements at the absorber can be approximated by:

\begin{eqnarray}
\langle x^2_{ab} \rangle & \simeq & R^2_{12}(s_{sp} \rightarrow s_{ab}) \phi^2_{MCS}\,\,, \label{absorbereqbetatron1}\\ 
\langle y^2_{ab} \rangle & \simeq & R^2_{34}(s_{sp} \rightarrow s_{ab}) \phi^2_{MCS}\,\,, \label{absorbereqbetatron2}
\end{eqnarray}

\noindent where $\phi_{MCS}$ is the angular divergence given by MCS in the spoiler. $R_{12}$ and $R_{34}$ are the transfer matrix elements between the betatron spoiler and the betatron absorber. In this case, $R_{12}(s_{sp} \rightarrow s_{ab})=114.04$~m and $R_{34}(s_{sp} \rightarrow s_{ab})=-483.22$~m from YSP1 to YAB1 (see spoiler names in Table~\ref{tablecoll5}). Taking into account $\sigma_x (s_{ab})=\sqrt{\langle x^2_{ab} \rangle}$ and $\sigma_y (s_{ab})=\sqrt{\langle y^2_{ab} \rangle}$, and Eqs.~(\ref{absorbereqbetatron1}) and (\ref{absorbereqbetatron2}), the condition for the survival of the betatron absorber can be written as follows:

\begin{equation}
\sqrt{|R_{12}(s_{sp}\rightarrow s_{ab})||R_{34}(s_{sp}\rightarrow s_{ab})|} \phi_{MCS} \gtrsim 600~\mu\textrm{m}\,\,,
\label{absorbereqbetatron3}
\end{equation}

\noindent which is fulfilled if $\phi_{MCS} \gtrsim 3 \times 10^{-6}$ rad. From this constraint and using the Gaussian approximation of the Moli\`ere formula for the MCS angle \cite{PDG}:

\begin{equation}
\phi_{MCS}=\frac{13.6~[\textrm{MeV}]}{\beta c p} z \sqrt{\frac{\ell}{X_0}} \left[ 1 + 0.038 \ln \left( \frac{\ell}{X_0}\right)\right] \,\, ,
\label{Coulombscatter}
\end{equation}

\noindent we can calculate the minimum length of spoiler material seen by an incident beam in order to guarantee the absorber survival. This condition is fulfilled if the Be spoiler is designed with a centre flat body of length $L_F \gtrsim 0.1~X_0$. For instance, selecting a spoiler with $L_F = 0.2~X_0$ could give a safe margin of angle divergence by MCS for absorber survival in case of beam impact.

Concerning the betatron spoiler protection for CLIC, it is worth mentioning that while the survival condition is important for the energy spoiler (see Sections 2.1.2 and 2.1.3 of Part I), it is not restrictive for the betatron spoilers. These spoilers are planned to be sacrificial, i.e. they would certainly be destroyed if they suffer the direct impact of a bunch train. Direct impacts on the betatron spoilers are expected to be infrequent events. Large betatron oscillations of on-energy beams are not easily generated from pulse to pulse, and in the linac they rapidly filament and emittance can increase by 2 orders of magnitude.   

In the hypothetical case that survivability of the betatron spoilers is desired, the betatron functions at the spoilers would have to be increased in order to enlarge the beam spot size sufficiently to ensure the spoiler survival. Nevertheless, this would increase the chromaticity of the lattice and generate tighter tolerances. 

%consequently, increase the beam spot size up to a value enough to ensure the spoiler survival if impacted by a bunch train.    
%Based on the SLC experience \footnote{The Stanford Linear Collider (SLC) \cite{SLC} is the sole linear collider built to date.}, energy errors in the linac are expected to occur much more frequently than orbit disruptions of on-energy beams. Large betatron oscillations are not easily generated from pulse to pulse, and in the linac they rapidly filament and emittance can increase by 2 orders of magnitude. Therefore,  

For CLIC the betatron spoilers have always been assumed to be made of Be. The main arguments to select Be were its high thermal and mechanical robustness and good electrical conductivity (to minimise resistive wakefields). Nevertheless, an important inconvenience of using Be is that its manipulation presents important technical challenges due to the toxicity of Be-containing dusts. An accident involving Be  might be a serious hazard. Since no survivability to the full beam power is demanded for the betatron spoilers, the robustness condition of the material could be relaxed, and different options other than Be could be investigated, e.g. Ti with Cu coating. 

%Spoiler designs based on the combination of different materials have been made for ILC \cite{spoilerILC1, spoilerILC2} and their properties have been studied by means of simulations \cite{spoilerILC1}, and by means of test beam studies at SLAC End Station A (ESA) \cite{spoilerILC4, spoilerILC5, spoilerILC6}. In principle, we could benefit from the experience gained from the ILC collimator design and adapt it to the CLIC requirements. 
If we decide to select a Ti based spoiler for betatron collimation, then, for absorber protection, the condition (\ref{absorbereqbetatron3}) is fulfilled if the spoiler is designed with a centre flat body (made of Ti) of length $L_F=0.2~X_0 \simeq 7$~mm.

Other proposals, such as rotating consumable collimators\footnote{Rotatable collimators are currently being constructed for the collimation upgrade of the Large Hadron Collider (LHC) \cite{rotableLHC}. The LHC collimation experience will be useful to guide the technical design, construction and upgrade of the CLIC collimators.} \cite{rotatingspoiler} and dielectric materials \cite{Kanareykin}, are being investigated as alternative for future upgrades of the design. 

Table~\ref{tablecoll6} shows the design parameters of the CLIC betatron spoilers and absorbers (of the baseline system) after optimisation.  

\begin{table}[!htb]
  \caption{Design parameters of the CLIC betatronic spoiler and absorbers.}
  \label{tablecoll6}
  \begin{center}
        \begin{tabular}{l c c}
       \hline \hline 
          \multicolumn{3}{c}{\bf Spoilers} \\
          \hline 
          Parameter & XSP$\#$ & YSP$\#$ \\      
          \hline 
          Geometry & Rectangular & Rectangular \\
          Hor. half-gap $a_x$ [mm] &  0.12 & 8.0 \\
          Vert. half-gap $a_y$ [mm]  &  8.0  & 0.1 \\
          Tapered part radius b [mm] &  8.0 & 8.0 \\
          Tapered part length $L_T$ [mm] & 90.0 & 90.0 \\ 
          Taper angle $\theta_T$ [mrad] &  88.0 & 88.0 \\
          Flat part length $L_F$ [radiation length] & 0.2 & 0.2  \\
          Material (other options?) & Be (Ti--Cu coating?) & Be (Ti--Cu coating?) \\                      
    \hline   
    \multicolumn{3}{c}{\bf Absorbers} \\
     \hline
          Parameter &  XAB$\#$ & YAB$\#$ \\      
          \hline 
          Geometry  & Circular & Circular \\
          Hor. half-gap $a_x$ [mm] & 1.0 & 1.0 \\
          Vert. half-gap $a_y$ [mm]  & 1.0 & 1.0 \\
          Tapered part radius b [mm] & 8.0 & 8.0 \\
          Tapered part length $L_T$ [mm] & 27.0 & 27.0 \\ 
          Taper angle $\theta_T$ [mrad] & 250.0 & 250.0 \\
          Flat part length $L_F$ [radiation length] & 18.0 & 18.0 \\
          Material & Ti alloy--Cu coating & Ti alloy--Cu coating \\
    
    \hline
    \end{tabular}
  \end{center}
 \end{table}   

\subsection{Optics optimisation}

\begin{figure}[htb]
\begin{center}
\includegraphics*[width=12cm]{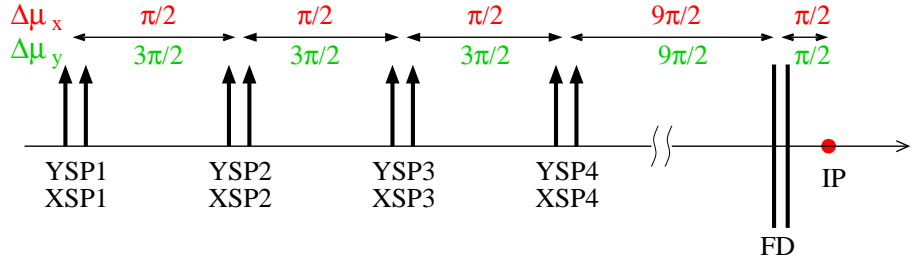}
\caption{Schematic showing the design values of the phase advance between the betatron spoilers, FD and IP.}
\label{phaseadvance}
\end{center}
\end{figure}

By design the phase advance of the betatron spoilers respect to the FD and the IP has to be matched to allow an efficient collimation of the transverse halo. The transverse phase advance between the spoiler positions and the IP is generally set to be $n\pi$ or $(1/2 + n)\pi$, with $n$ an integer. Figure~\ref{phaseadvance} illustrates the design transverse phase advances of the CLIC betatron spoilers. The IP is at $\pi/2$ phase advance from the FD, and the phase relationship between the betatron collimators and the FD is crucial. The spoilers XSP1 (YSP1) and XSP3 (YSP3) are set to collimate amplitudes at the FD phase, while the spoilers XSP2 (YSP2) and XSP4 (YSP4) collimate amplitudes at the IP phase. 

In the CLIC lattice version 2008 the phase advances between the fourth set of spoilers (YSP4 and XSP4) and the FD were not an exact multiple of $\pi/2$: $\Delta \mu^{SP4 \rightarrow FD}_{x,y} = 9.7 \pi/2,10.6 \pi/2 $. Starting from this original lattice, and following a similar phase optimisation procedure as it was used for the ILC \cite{Jackson1, Jackson2}, we have investigated phase-matched solutions between the fourth set of spoilers and the FD in order to further improve the collimation performance of the system. In this study the software MAD \cite{MAD} has been used to model the lattice and perform the phase matching. In total eight quadrupoles have been used for the matching: four of them (BTFQ1, BTFQ2, BTFQ3 and BTFQ4) at the end of the betatron collimation section (Fig.~\ref{latticebetatroncoll}) and four quadrupoles (QMD11, QMD12, QMD13 and QMD14) at the beginning of the FFS. Here the quadrupoles are named as in the CLIC lattice repository of Ref.~\cite{CLICrepository} 

\begin{figure}[htb]
\begin{center}
\includegraphics*[width=7cm]{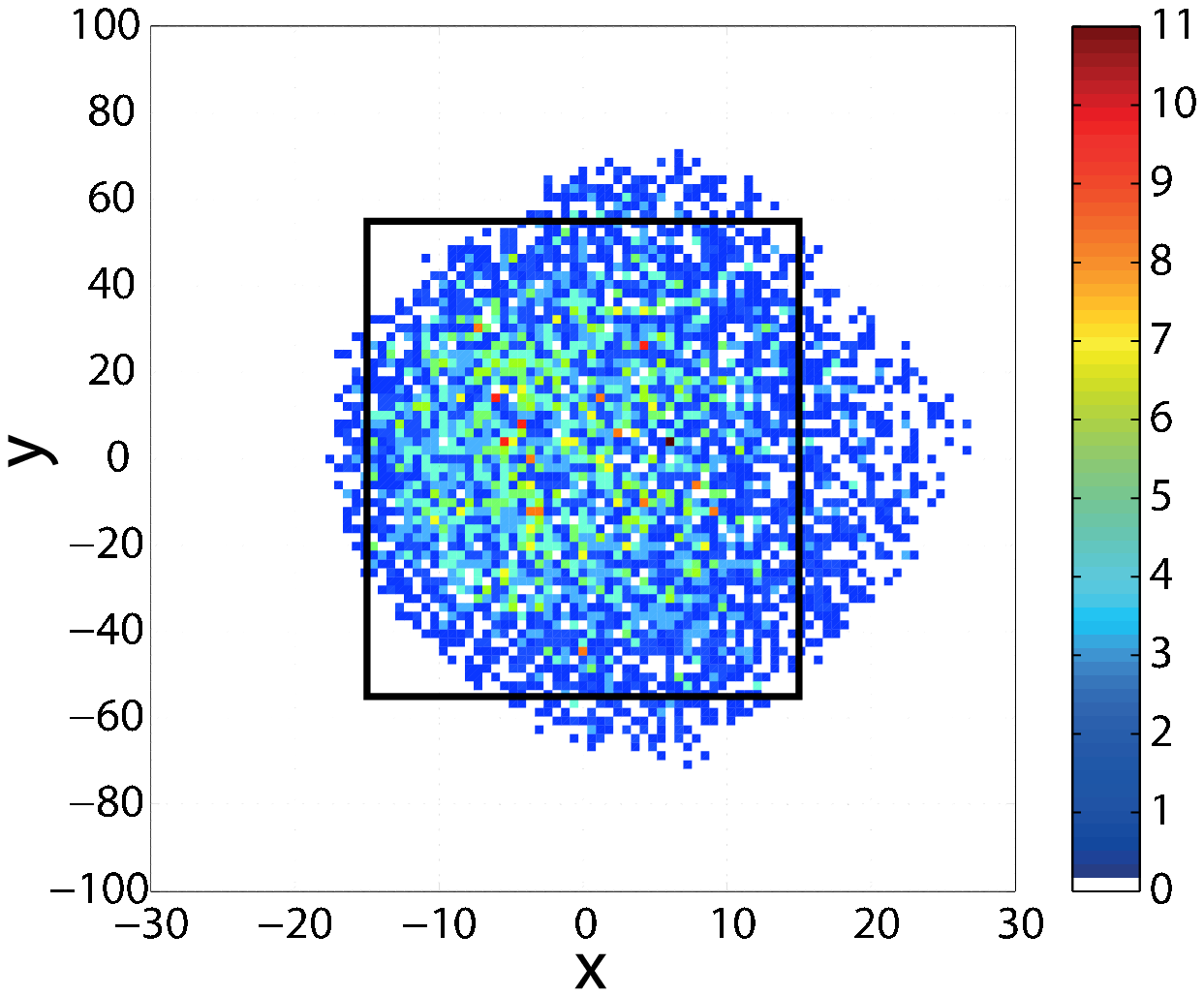}
\includegraphics*[width=7cm]{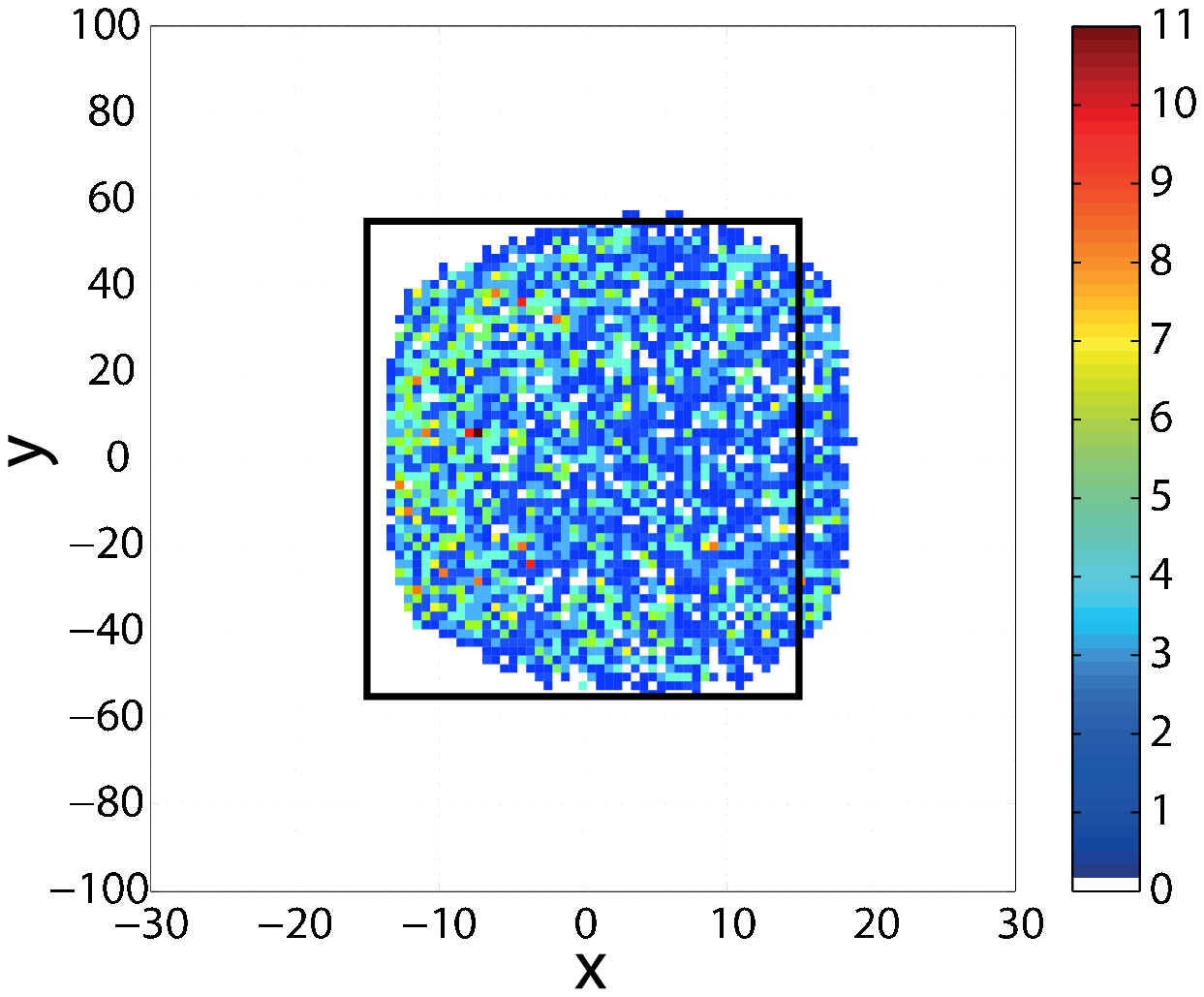}
\caption{Beam halo $x$--$y$ profile at the entrance of the FD (Left) and after (Right) optimisation. In this example no beam energy spread has been considered. The black square contour represents the collimation window.}
\label{FDprofile}
\end{center}
\end{figure}

The collimation performance of the lattices has been evaluated from beam halo tracking simulations using the code MERLIN \cite{MERLIN}. For the tracking a ``toy'' model of the primary beam halo, consisting of 25000 macroparticles with energy 1500~GeV and zero energy spread, was generated at the BDS entrance, uniformly distributed in the phase spaces $x$--$x'$ and $y$--$y'$ and extending to 1.5 times the collimation depth. The halo has been tracked from the BDS entrance to the IP, treating the collimators as perfect absorbers of any incident particle. A measure of the primary collimation efficiency is the number of particles outside the collimation depth at the FD. A phase-matched solution has been found at $\Delta \mu^{SP4 \rightarrow FD}_{x,y} = 10 \pi/2,11 \pi/2$, which reduces the ``escaped particles'' (outside the collimation window) by $20\%$ with respect to the original lattice. The strength values of the matching quadrupoles of the optimised lattice are shown in Table~\ref{quadstrengths}, compared with the initial values of the original lattice. The pole tip radius aperture for these quadrupoles is 8~mm. The effective lengths of the quadrupoles are 5~m for the BTFQ$\#$ type quadrupole and $1.63$~m for QMD$\#$. Figure~\ref{FDprofile} compares the halo $x$--$y$ profile at the FD entrance for the original and the new matched lattices.  

\begin{table}[!htb]
  \caption{Strength of matching quadrupoles in the transition region between the betatron collimation section and the final focus system for the original lattice and for the optimised lattice. $K$ and $B_0$ denote the integrated quadrupole strength and the pole tip magnetic field, respectively.}
  \label{quadstrengths}
  \begin{center}
        \begin{tabular}{|l|c|c||c|c|}
       \hline \hline 
          {} & \multicolumn{2}{c||}{Original} & \multicolumn{2}{c|}{Optimisation} \\
          \hline 
          Name & $K$~[m$^{-1}$] & $B_0$~[T] & $K$~[m$^{-1}$] & $B_0$~[T] \\      
          \hline \hline
          BTFQ1 & -0.0605 & 0.48432 & -0.0669 & 0.5356 \\
          BTFQ2 & 0.0152 & 0.1217 & 0.0386 & 0.309 \\
          BTFQ3 & 0.0252 & 0.2017 & 0.0285 & 0.2281 \\
          BTFQ4 & -0.0333 & 0.2666 & -0.0731 & 0.5852 \\
          QMD11 & 0.0905 & 2.2224 & 0.1551 & 3.8087 \\
          QMD12 & -0.1423 & 3.4944 & -0.1023 & 2.5121 \\
          QMD13 & 0.1095 & 2.6889 & 0.0961 & 2.3599 \\
          QMD14 & -0.0502 & 1.2327 & -0.0736 & 1.8074 \\   
    \hline   
\end{tabular}
\end{center}
\end{table}

In addition to the collimation optimisation, it is necessary to evaluate the impact of the lattice changes on the luminosity. It is important that the lattices optimised for collimation maintain good properties for beam luminosity. We have calculated the luminosity peak from beam-beam interaction simulations at the IP by using the code GUINEA-PIG \cite{guinea} for both the original and the optimised lattices. For the tracking of a Gaussian core beam with a uniform energy distribution of $1~\%$ full energy spread, results show that the luminosity peak of the optimised lattice (for betatron collimation) is 1.5 times smaller than the luminosity of the original lattice. In principle, the luminosity performance can be recovered by adjusting slightly the strengths of the final focus sextupoles as it was made for the ILC BDS optimisation \cite{Jackson2}. For these tracking simulations and luminosity calculations the CLIC BDS lattice of $L^{*}=3.5$~m (available at the CLIC lattice repository \cite{CLICrepository}) has been used, where $L^{*}$ is the distance from the final quadrupole QD0 to the IP.     

\section{Wakefield effects}
\label{wakefieldsection}

A charged particle moving in an accelerator induces electromagnetic fields which interact with its environment. Depending on the discontinuities and variations in the cross-sectional shape of the vacuum chamber, the beam self field is perturbed and can be reflected onto the beam axis and interact with particles in the beam itself. These electromagnetic fields, induced by the charged beam, are called wakefields, due to the fact that they are left mainly behind the driving charge (the source charge of the wakefield). In the limit of ultra-relativistic motion the wakefields can only stay behind the driving charge. 

In the case of bunched beams, depending on whether the wakefields interact with the driving bunch itself or with the following bunches, they are denominated \emph{short range wakefields} or \emph{long range wakefields}, respectively. The former may degrade the longitudinal and transverse emittances of individual bunches and the latter may cause collective instabilities.

Wakefields in the BDS of the linear colliders can be an important source of emittance growth and beam jitter amplification, consequently degrading the luminosity. The main contributions to wakefields in the BDS are:

\begin{itemize}

\item Geometric and resistive wall wakefields of the tapered and flat parts of the collimators. 

\item Resistive wall wakes of the beam pipe, which are especially important in the regions of the final quadrupoles, where the betatron functions are very large.

\item Electromagnetic modes induced in crab cavities. Crab cavities are needed to rotate the train bunches in order to compensate for the crossing angle at the IP, which is $20$~mrad in the case of CLIC. 

\end{itemize} 

\noindent In this report we focus on single bunch effects of the collimator transverse wakefields.

The main contribution to the collimator wakefields arises from the betatron spoilers, whose apertures ($\approx 100~\mu$m) are much smaller than the design aperture of the energy spoiler ($3.5$~mm), and much smaller than the aperture of the nearby vacuum chamber (8--10~mm radius).

\begin{figure}[htb]
\begin{center}
\includegraphics*[width=12cm]{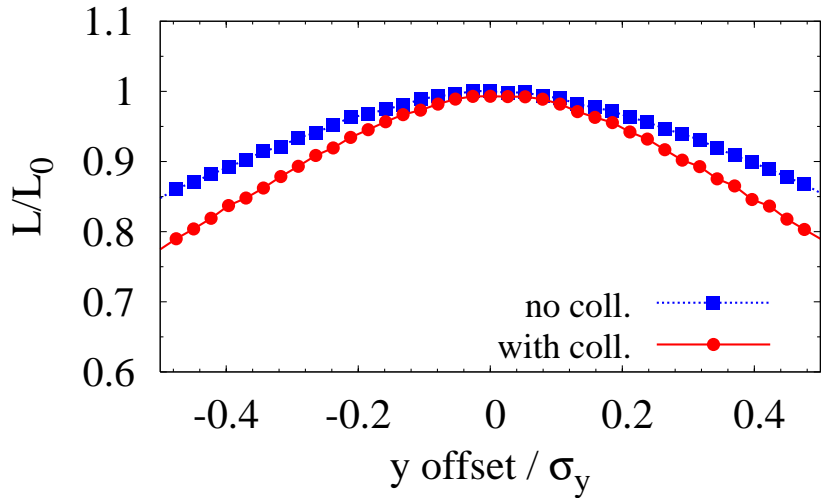}
\caption{Relative CLIC luminosity versus initial beam offsets for the cases with and without collimator wakefield effects.}
\label{wakeeffects1}
\end{center}
\end{figure}

In order to study the impact of the CLIC collimator wakefields on the beam, a module for the calculation of the collimator wakefields in different regimes has been implemented in the PLACET tracking code \cite{Giovanni}. Using this code the effects of the collimator wakefields on the luminosity have been evaluated for the design transverse collimation apertures $15~\sigma_x$ and $55~\sigma_y$. Figure~\ref{wakeeffects1} compares the relative luminosity degradation as a function of initial vertical position offsets at the entrance of the BDS with and without collimator wakefields. In this calculation the join effect of all the BDS collimators has been considered. For instance, for beam offsets of $\approx \pm 0.4~\sigma_y$, the CLIC luminosity loss was found to amount up to $20\%$ with collimator wakefields, and up to $10\%$ for the case with no wakefield effects.

\begin{figure}[htb]
\begin{center}
\includegraphics*[width=12cm]{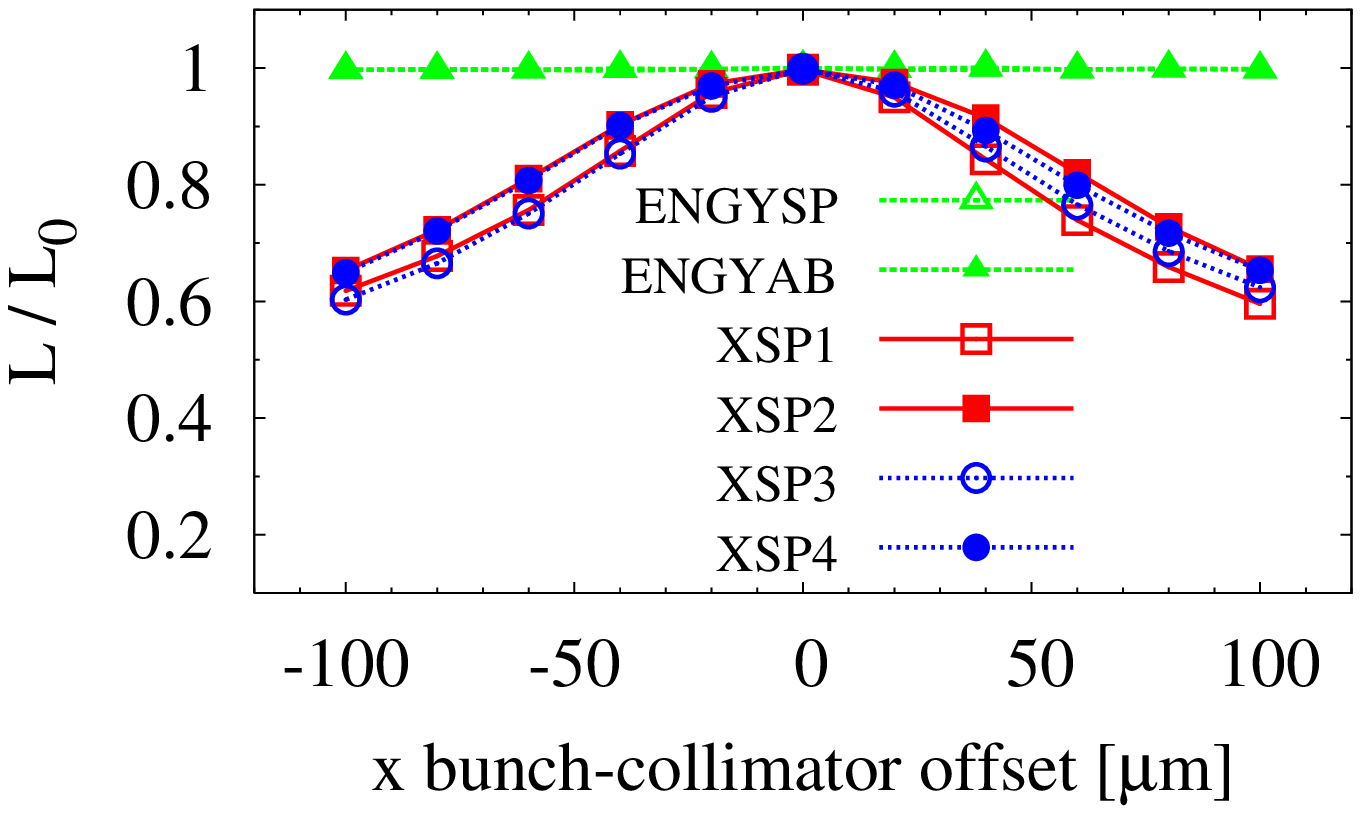}
\includegraphics*[width=12cm]{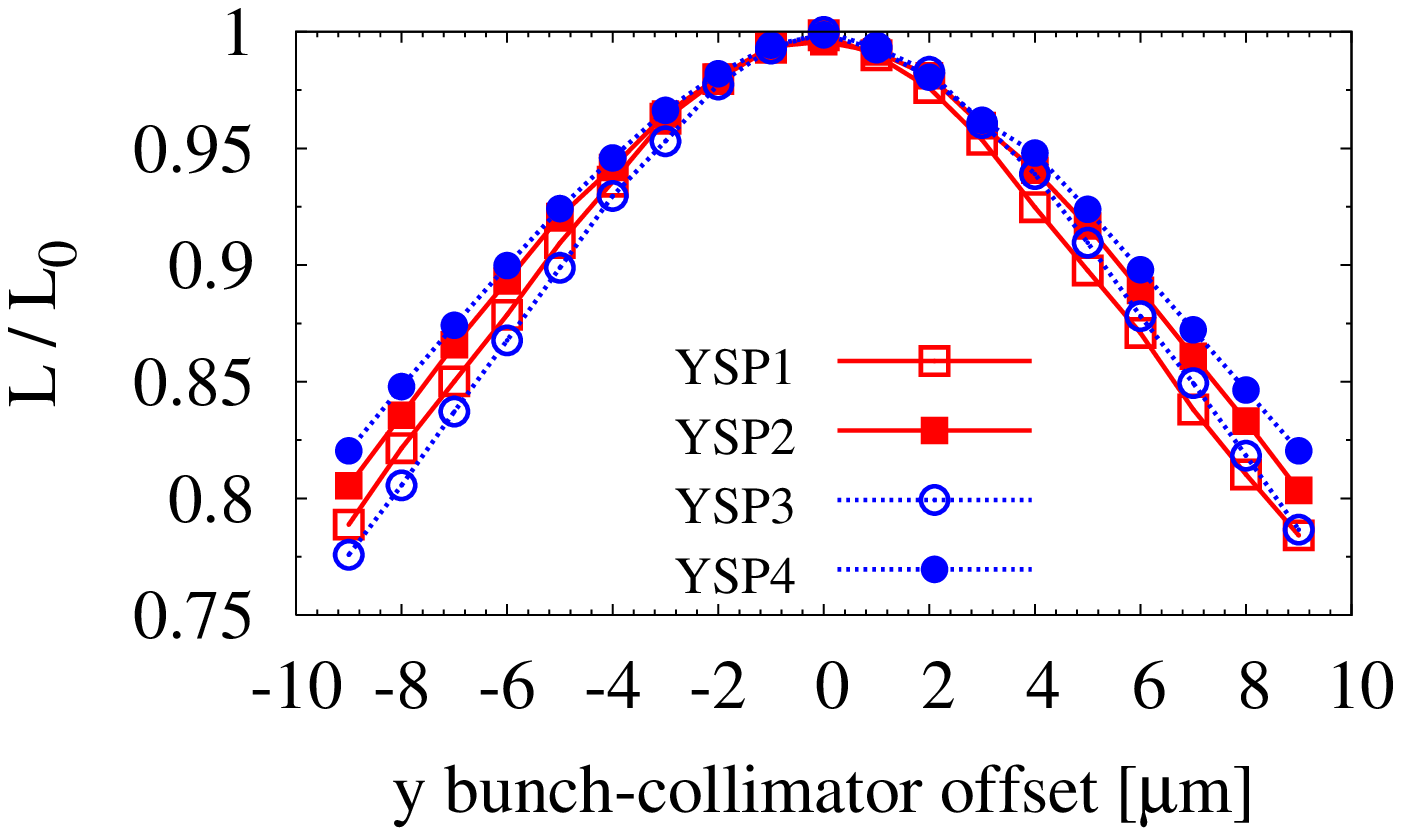}
\caption{Top: relative luminosity versus horizontal bunch-collimator offset for each rectangular horizontal collimator. Bottom: relative luminosity versus vertical bunch-collimator offset for each rectangular vertical collimator.}
\label{wakeeffects2}
\end{center}
\end{figure}

The luminosity loss due to horizontal misalignments (with respect to the on-axis beam) of each horizontal spoiler and absorber is shown in Fig.~\ref{wakeeffects2} (Top). In comparison with the betatron collimators the energy spoiler (ENGYSP) and the energy absorber (ENGYAB) have been set with a large half gap, and practically do not contribute to the luminosity degradation by wakefields. On the other hand, for the horizontal betatron spoilers $\approx 20\%$ luminosity loss is obtained for $\approx \pm 50~\mu$m bunch-collimator offset. 

In the same way, Fig.~\ref{wakeeffects2} (Bottom) shows the relative luminosity as a function of the vertical bunch-collimator offset for each vertical betatron spoiler. The stronger wake kick effects arise from the spoilers YSP1 and YSP3. Approximately $20\%$ luminosity loss is obtained for vertical bunch-collimator offsets of $\approx \pm 8~\mu$m.   

%the luminosity loss due to vertical misalignments of the vertical spoilers is shown in Fig.~\ref{wakeeffects2} (Bottom). Approximately $20\%$ luminosity loss is obtained for $\approx \pm 9~\mu$m bunch-collimator offset.

%The beam stability in the $y$--$y'$ phase space is more critical for the luminosity performance, since it is more sensitive to errors than the horizontal one due to the transverse flatness of the beam ($\sigma^{*}_y/\sigma^{*}_x \approx 1/100$)

%More critical for the luminosity performance is 

%jaws are much closer to the beam than those of the energy spoiler. 

%Wakefield effects implemented in the code PLACET-Octave \cite{Giovanni}. 

In order to optimise the spoiler design and thus reduce the wakefield effects, the following items could be investigated:

\begin{itemize}

\item Decreasing the geometrical wakes by optimising the spoiler taper angle.

\item Coating the main body of the spoiler with a very thin layer of a very good electrical conductor.  

\item Exploring novel concepts, e.g. dielectric collimators \cite{Kanareykin}.

\end{itemize}

\subsection{Spoiler taper angle optimisation}

Let us consider a beam with centroid offset $y_0$ from the beam axis passing through a symmetric spoiler of minimum half gap $a$. Assuming $y_0 \ll a$, the mean beam deflection due to spoiler wakefields can be  expressed as follows:

\begin{equation}
\langle y' \rangle = \frac{r_e N_e}{\gamma} \kappa y_0 \,\,,
\label{kickanglewake}
\end{equation}   

\noindent where $r_e$ is the electron classical radius, $N_e$ the number of particles per bunch and $\gamma \equiv E/(m_e c^2)$ the relativistic Lorentz factor, with $E$ the beam energy, $m_e$ the rest mass of the electron and $c$ the speed of light. In this equation the beam deflection has been given in terms of a transverse wake kick factor  $\kappa=\kappa_g + \kappa_r$, which can be expressed as the sum of a geometrical wake kick contribution, $\kappa_g$, and another kick factor taking into account the resistive wall contribution,  $\kappa_r$.

The spoilers are commonly designed with shallow taper angles in order to reduce the geometrical component of the wakefields. The taper angle is 88 mrad in the current design of the betatron spoiler (see Table~\ref{tablecoll6}). Here we investigate  the possibility of reducing the wakefield effects by optimising the taper angle of the spoilers and, in consequence, to improve the luminosity performance. 

For the taper angle optimisation we have to take into account the different collimator wakefield regimes as the taper angle changes. The geometrical wake kick can be calculated using the following ``near-centre'' approximation for rectangular collimators \cite{Stupakov}: 

\begin{equation}
\footnotesize{
\kappa_g = \left\{ \begin{array}{lll}
\sqrt{\pi} \theta_{T} h/(2\sigma_z) \left(1/a^2 - 1/b^2 \right) & \textrm{for} & \theta_T < 3.1^2 a \sigma_z/h^2 \,\,, \\
8/3 \sqrt{\theta_T/(\sigma_z a^3)} & \textrm{for} & 0.37^2 \sigma_z/a > \theta_T > 3.1^2 a \sigma_z/h^2 \,\,, \\
1/a^2 & \textrm{for} & \theta_T > 0.37^2\sigma_z/a \,\,. 
\end{array} \right.
}
\label{kickgeom1}
\end{equation} 

\noindent As before, $b$ and $a$ denote the maximum and minimum half gap of the  collimator, respectively. Here, $h$ denotes the half width of the gap in the non-collimating direction. In Eq.~(\ref{kickgeom1}) the limit $ \theta_T < 3.1^2 a \sigma_z/h^2$ corresponds to the inductive regime; $0.37^2 \sigma_z/a > \theta_T > 3.1^2 a \sigma_z/h^2$ corresponds to the intermediate regime; and $\theta_T > 0.37^2\sigma_z/a$ the diffractive regime. Considering the parameters for the vertical betatron spoiler of CLIC (Table~\ref{tablecoll6}), Eq.~(\ref{kickgeom1}) can be written as follows: 

\begin{equation}
\footnotesize{
\kappa_g = \left\{ \begin{array}{lll}
\sqrt{\pi} \theta_{T} h /(2\sigma_z) \left(1/a^2 - 1/b^2 \right) & \textrm{for} & \theta_T < 7 \times 10^{-4}~\textrm{rad}\,\,, \\
8/3 \sqrt{\theta_T/(\sigma_z a^3)} & \textrm{for} & 0.06~\textrm{rad} > \theta_T > 7 \times 10^{-4}~\textrm{rad}\,\,, \\
1/a^2 & \textrm{for} &  \theta_T > 0.06~\textrm{rad}\,\,. 
\end{array} \right.
}
\label{kickgeom2}
\end{equation} 

\noindent For flat rectangular tapered spoilers the kick factor corresponding to the resistive component of the collimator wakefield  can be approximate by the following expression for very small beam offsets \cite{resistivekick}:

\begin{equation}
\kappa_r \simeq \frac{\pi}{8a^2} \Gamma(1/4)\sqrt{\frac{2}{\sigma_z \sigma Z_0}} \left[ \frac{L_F}{a} + \frac{1}{\theta_T} \right] \,\,,
\label{kickres}
\end{equation}  

\noindent where $Z_0=376.7~\Omega$ is the impedance of free space and $\Gamma(1/4)=3.6256$. 

\begin{figure}[htb]
\begin{center}
\includegraphics*[width=8cm, angle=-90]{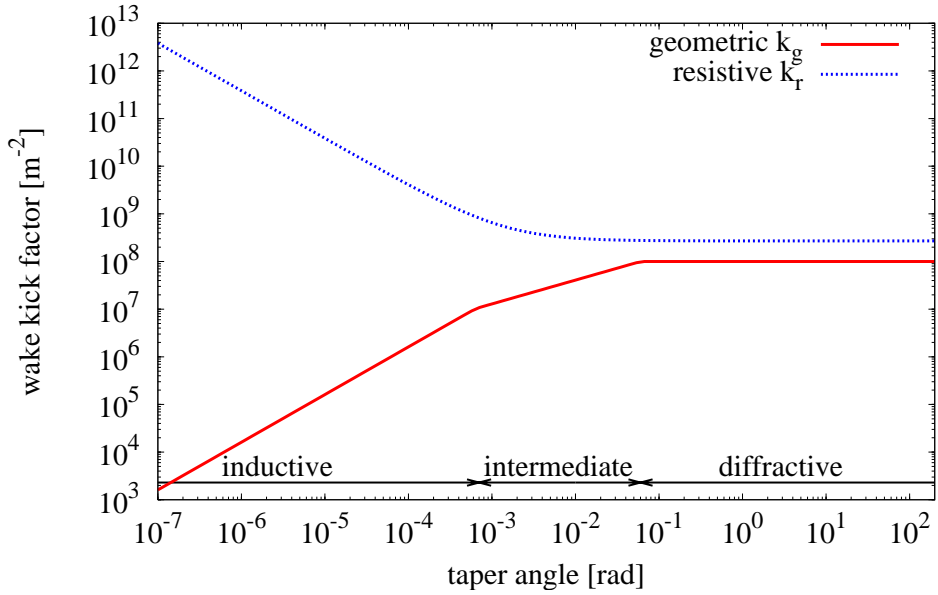}
\caption{Geometric and resistive wake kick factors versus spoiler taper angle calculated from Eq.~(\ref{kickgeom2}) and Eq.~(\ref{kickres}), respectively. The $x$--axis and $y$--axis are on logarithmic scale. The different regimes for the geometric wakefields are indicated.}
\label{plotwakekick}
\end{center}
\end{figure}

\noindent The wake kick generated by a CLIC betatron spoiler in the vertical plane as a function of the taper angle is represented in Fig.~\ref{plotwakekick}, where the geometric and the resistive contribution are shown separately. With taper angle 88 mrad the geometric kick is in the diffractive regime. One could expect to reduce the geometric wakes by reducing the taper angle. However, on the other hand, for CLIC the resistive wake kick is dominant, and it increases as $1/\theta_T$ as the taper angle is decreased.  

%For CLIC the resistive kick is dominant, and it increases as $1/\theta_T$ as the taper angle is decreased. 

\begin{figure}[htb]
\begin{center}
\includegraphics*[width=7.5cm, angle=-90]{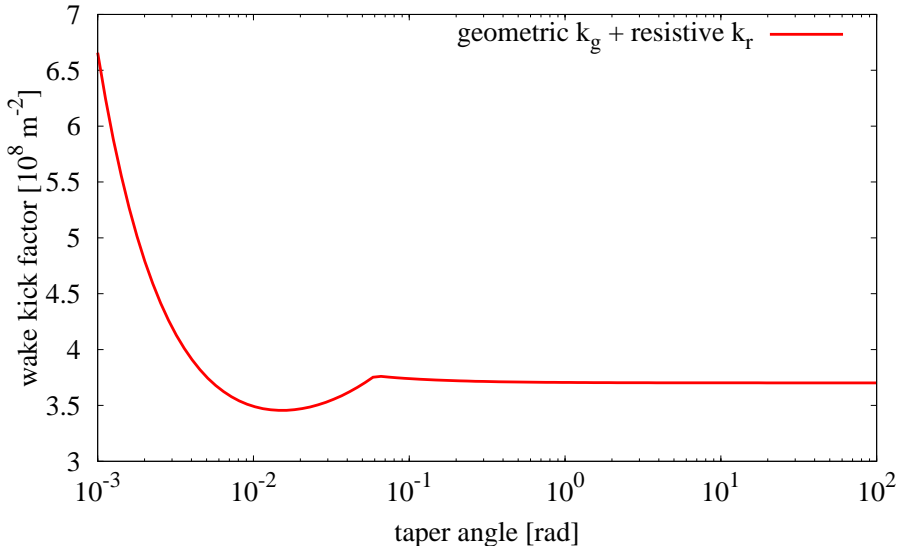}
\caption{Total wake kick factor versus spoiler taper angle. The $x$--axis is on a logarithmic scale.}
\label{plotwakekicktotal}
\end{center}
\end{figure}

The total wake kick, adding both geometric and resistive contributions, is shown in Fig.~\ref{plotwakekicktotal}. For taper angles $< 0.01$ rad the total wake kick strongly increases due to the resistive wake dominance. For angles $> 0.1$ rad the wake kick is not very sensitive to the change in the taper angle and remains practically constant. In Fig.~\ref{plotwakekicktotal} one can also note that there is a minimum wake kick factor in between 0.01 and 0.02 rad. For example, in order to improve the performance of the system in terms of wakefields, a new taper angle of $\approx 15$ mrad could be selected. However, doing this, it is also necessary to increase the total longitudinal length of the spoiler, $2 L_T + L_F$, from 25 cm (for the original taper angle 88 mrad) to $\approx 1$ m for the new taper angle. Therefore, decreasing the taper angle one has to deal with a longer spoiler, and, given the tiny aperture of the betatron spoilers, tilt errors in the spoiler alignment could have much more negative effects on the beam stability than those affecting shorter spoiler.          

% Explicar the en este plot the resistive contribution solo incluye the taper contribution (proporcional a $1/\theta_T$).  

\subsection{Betatron spoiler design review with regard to wakefields}

In previous sections both energy and betatron spoilers have been considered made of Be. Beryllium was selected due to its high thermo-mechanical robustness as well as its high electrical conductivity in comparison with other metals. However, due to the highly toxicity of Be dust, special care must be taken when machining the material.

Since the betatronic spoilers are not required to survive the impact of an entire bunch train, i.e. they are planned to be sacrificial, in principle we could investigate optional materials other than Be for the betatronic spoiler design. Preliminary studies of spoiler design options with different geometry and combining different metals were shown in \cite{RestaandLuis, LuisandResta}. For example, Ti alloy ($90\%$ Ti, $6\%$ Al, $4\%$ V) and Ti alloy with Cu coating could be good alternatives for the betatronic spoilers. 

Be is better conductor than Ti: the electrical conductivity of Be at room temperature ($\sigma (\textrm{Be}) \simeq 2.3 \times 10^7$~$\Omega^{-1}$m$^{-1}$) is one order of magnitude higher than the Ti conductivity ($\sigma (\textrm{Ti}) \simeq 1.8 \times 10^6$~$\Omega^{-1}$m$^{-1}$). Therefore, in terms of wakefields Be is a better option than Ti. As we have seen in previous sections, for the current design of the CLIC spoilers, the main contribution to the wakefields is basically resistive. From Eq.~(\ref{kickres}) the dependence of the resistive wakefield kick on the electrical conductivity ($\sigma$) is given by $\kappa_r \propto 1/\sqrt{\sigma}$. The resistive wakefield kick by a Ti spoiler is almost four times bigger than the kick by a Be spoiler, $\kappa_r (Ti)/\kappa_r (Be) = \sqrt{\sigma (Be)}/\sqrt{\sigma (Ti)} \simeq 4$. On the other hand, the resistive kick produced by a Be spoiler is almost two times bigger than the kick by a spoiler made of Cu, $\kappa_r (Be)/\kappa_r (Cu) = \sqrt{\sigma (Cu)}/\sqrt{\sigma (Be)} \simeq 2$. Betatronic spoilers made of Ti coated with Cu could be a good option to reduce the impact of wakefields. 

Other line of investigation, aimed to minimise the collimator wakefields, has recently started the design of dielectric collimators for the CLIC BDS \cite{Kanareykin}. Dielectric collimators are currently being designed for the second phase of collimation of the Large Hadron Collider (LHC) \cite{Metral}. The plan is to adapt this concept also to the CLIC requirements. In \cite{Kanareykin} preliminary wakefield calculations have been made considering a cylindrical geometry model consisting of double layer based on a dielectric material coated with an external layer of copper.         

\subsection{Deflection due to surface roughness}
\label{surfaceroghness}

As seen in Section~2.1.3 (Part I), the impact of the full beam onto the Be spoiler might cause a permanent deformation of the spoiler surface. This could increase the wakefield effects and, therefore, to have negative consequences on the beam stability.  

The average kick angle due to wakefield effects caused by the roughness of the spoiler/collimator surface can be estimated using the following expression for tapered surfaces \cite{PT2}:

\begin{equation}
\langle x' \rangle_{\rm rough} \simeq \frac{4}{3 a^2 \pi^{3/2}} \frac{N_e r_e}{\gamma \sigma_z} \left[ \frac{L_F}{a} + \frac{1}{\theta_T} \right] \zeta f \alpha_s x_0 \,\,,
\label{roughnesseq}
\end{equation}

\noindent where $\zeta$ is the characteristic size of the feature caused by the deformation, $f$ is a form factor for the shape of the features, which is typically in the range between 1 and 20, $\alpha_s$ is the fraction of the surface filled with the features, $N_e$ is the bunch population, $r_e$ the electron classical radius, and $x_0$ the offset of the beam centroid with respect to the nominal beam axis. As in previous sections, $a$ and $b$ denote the minimum and maximum spoiler half gap, respectively. While in Ref.~\cite{PT2} only the tapered contribution ($1/\theta_T$) was taken into account, here Eq.~(\ref{roughnesseq}) takes into account the contributions from the tapered part and from the flat part ($L_F/a$).     

According the ANSYS results of Section~2.1.3 (Part~I), horizontal deformation protuberances of about $\zeta \approx 1~\mu$m might be caused by tensile stress in the E-spoiler. We can roughly estimate the angular deflection using Eq.~(\ref{roughnesseq}). For one hemispherical bump, the form factor $f=3\pi/2$. For example, if we assume $\alpha_s \approx 1/3$ and $x_0 \approx 1~\sigma_x$ (with $\sigma_x = 779~\mu$m at the E-spoiler), we obtain $\langle x' \rangle_{\rm rough} \simeq 4.8 \times 10^{-11}$~rad, which is approximately a factor 3 larger than the resistive wakefield kick $\langle x' \rangle_{\rm resistive} \simeq 1.8 \times 10^{-11}$~rad obtained from Eq.~(\ref{kickres}) for the same beam offset $x_0 \approx 1~\sigma_x$ and for the E-spoiler. 

If now we assume the same hypothetical level of deformation in a CLIC vertical betatronic spoiler made of Be, according Eq.~(\ref{roughnesseq}), one obtains $\langle y' \rangle_{\rm rough} \simeq 3.6 \times 10^{-9}$~rad for a beam vertical deviation of $10~\sigma_y$ (with $\sigma_y= 1.814~\mu$m). This value is approximately $24\%$ of the value obtained for the resistive wake kick, $\langle y' \rangle_{\rm resistive} \simeq 1.5 \times 10^{-8}$ for the same vertical beam offset.

\section{Collimation system for CLIC at 500 GeV CM energy}

The optics design of the CLIC BDS for 500 GeV CM energy can be found in the CLIC lattice repository of Ref.~\cite{CLICrepository}, where it is available in the format of the codes PLACET \cite{placetoctave} and MAD \cite{MAD}. For this energy option the collimation section is almost two times shorter than that of CLIC at 3~TeV. In total, the CLIC BDS length ratio for the options 0.5~TeV/3~TeV is $1.73$~km/$2.79$~km. 

No optimisation of the collimation parameters has yet been made for this option. In principle, the same collimation depths as well as the same number of collimators have been assumed for both 500 GeV and 3~TeV. The betatron functions, horizontal dispersion and rms beam sizes at each collimator position for the 500 GeV case are shown in Table~\ref{optics500}. 

\begin{table}[!htb]
%\begin{sidewaystable}
\caption{Optics and beam parameters at collimator position for CLIC at 500 GeV CM energy: longitudinal position, horizontal and vertical $\beta$-functions, horizontal dispersion, horizontal and vertical rms beam sizes. ENGYSP and ENGYAB denote the energy spoiler and the energy absorber, respectively. SP$\#$ denotes vertical spoiler, XSP$\#$ horizontal spoiler, YAB$\#$ vertical absorber and XAB$\#$ horizontal absorber. The rms horizontal beam size at the energy collimators has been calculated assuming a uniform energy distribution with $1\%$ full energy spread.}
\label{optics500}
\begin{center}
\begin{tabular}{|l|c|c|c|c|c|c|}
\hline 
Name & s [m] & $\beta_{x}$ [m] & $\beta_{y}$ [m] & $D_x$ [m] & $\sigma_x$ [$\mu$m] & $\sigma_y$ [$\mu$m] \\
\hline \hline
ENGYSP & 453.549 & 703.166 & 35340.91 & 0.231 & 670.939 & 42.5 \\
ENGYAB & 536.049 & 1606.516 & 19635.742 & 0.357 & 1034.994 & 31.676 \\
YSP1 & 915.436 & 57.027 & 241.653 & 0. & 16.726 & 3.514 \\
XSP1 & 923.347 & 135.001 & 50.678 & 0. & 25.734 & 1.609 \\
XAB1 & 961.946 & 135.051 & 40.446 & 0. & 25.739 & 1.438 \\
YAB1 & 970.857 & 57.027 & 241.565 & 0. & 16.726 & 3.513 \\
YSP2 & 971.857 & 57.027 & 241.567 & 0. & 16.726 & 3.513 \\
XSP2 & 979.768 & 135.001 & 50.676 & 0. & 25.734 & 1.609 \\
XAB2 & 1018.368 & 135.052 & 40.478 & 0. & 25.739 & 1.438 \\
YAB2 & 1027.279 & 57.027 & 241.655 & 0. & 16.726 & 3.514 \\
YSP3 & 1028.279 & 57.027 & 241.653 & 0. & 16.726 & 3.514 \\
XSP3 & 1036.19 & 135.001 & 50.679 & 0. & 25.734 & 1.609 \\
XAB3 & 1074.789 & 135.051 & 40.446 & 0. & 25.739 & 1.438 \\
YAB3 & 1083.7 & 57.027 & 241.565 & 0. & 16.726 & 3.513 \\
YSP4 & 1084.7 & 57.027 & 241.568 & 0. & 16.726 &  3.513 \\
XSP4 & 1092.611 & 135.001 & 50.676 & 0. & 25.734 & 1.609\\ 
XAB4 & 1131.211 & 135.052 & 40.478 & 0. & 25.739 & 1.438 \\
YAB4 & 1140.122 & 57.027 & 241.655 & 0. & 16.726 & 3.514 \\
\hline 
\end{tabular}
\end{center}
%\end{sidewaystable}
\end{table}

% Explicar aqui las cosas de la radiacion sincrotron!!!!!!!!
For the CLIC optics at 500 GeV the dispersion $D_x$ at the energy spoiler and absorber positions has been decreased $\approx 14\%$ with respect to the 3 TeV optics. Taking into account that the emittance dilution due to incoherent synchrotron radiation scale as $\Delta(\gamma \epsilon_x)\propto E^6 D^5_x/L^5$, where $E$ is the beam energy and $L$ the total length of the collimation lattice, then for the 500 GeV case the relative emittance growth ($\Delta \epsilon_x / \epsilon_x$) in the collimation system is expected to be about four orders of magnitude smaller than for the 3 TeV case. Table~\ref{tableemittanceg} compares the horizontal emittances growth and luminosity loss for the 500 GeV and 3 TeV cases as calculated using Eqs.~(1) and (2) of Part I.

\begin{table}[!htb]
%\begin{sidewaystable}
\caption{Radiation integral $I_5$, relative emittance growth ($\Delta \epsilon_x / \epsilon_x$) and relative luminosity loss ($\Delta \mathcal{L} /\mathcal{L}$) due to synchrotron radiation in the collimation system and in the total BDS calculated for CLIC at 3~TeV and 0.5~TeV CM energy.}
\label{tableemittanceg}
\begin{center}
\begin{tabular}{|l|c|c||c|c|}
\hline \hline 
{} & \multicolumn{2}{c||}{\bf CLIC 3 TeV} & \multicolumn{2}{c|}{\bf CLIC 0.5 TeV} \\
\hline
{Variable} & Coll. system & Total BDS & Coll. system  & Total BDS \\
\hline \hline
$I_5$ [m$^{-1}$] & $1.9 \times 10^{-19}$ & $3.8 \times 10^{-19}$ & $5.6 \times 10^{-18}$ & $7.3 \times 10^{-16}$ \\
$\Delta \epsilon_x / \epsilon_x$ [$\%$] & 13.5 & 27.3 & 0.0023 & 0.31 \\
$\Delta \mathcal{L} /\mathcal{L}$ [$\%$] & 6.1 & 11.4 & 0.0012 & 0.15 \\
\hline
\end{tabular}
\end{center}
%\end{sidewaystable}
\end{table} 

Comparing the two energy cases, the following observations can be made:     

\begin{itemize}
\item For CLIC at 500 GeV the beam power is 4.8 MW, which is $\approx 66\%$ lower than that for CLIC at 3 TeV. Therefore, for CLIC at 500 GeV the damage potential of the beam (250 GeV beam energy) is smaller than that for the 3 TeV case (1.5 TeV energy beam), and more relaxed survival conditions can be considered for the energy spoiler. In this respect, materials with a lower fracture limit than Be may be chosen. A possible canditate might be Ti alloy.    

\item In order to minimise the multi-bunch effects of resistive wall in the CLIC BDS, the beam pipe radius was set at $b=10$~mm for the 3 TeV case. Since for the CLIC at 500 GeV the beam charge is higher, the beam pipe radius has been set at $b=12$~mm \cite{Mutzner}.   

\item Considering the same collimation depths $15~\sigma_x$ and $55~\sigma_y$, Table~\ref{apertures500} compares the collimator half gaps for both 500 GeV and 3 TeV options. 

\item The geometrical parameters of the collimators have to be calculated according the above minimum and maximum apertures. For instance, we can simply assume the same length for the collimators and then calculate the corresponding taper angles, $\theta_T=\tan^{-1} ((b-a)/L_T)$. 

\item In this preliminary design the collimators (spoilers and absorbers) have been assumed to be made of similar materials and with the same geometrical structure as described in Section~\ref{betatronsection}. 
\end{itemize}

\begin{table}[!htb]
%\begin{sidewaystable}
\caption{Half gaps of the CLIC post-linac collimators for the options at 3~TeV and 0.5~TeV CM energy. The values in parenthesis are new apertures suggested after optimisation.}
\label{apertures500}
\begin{center}
\begin{tabular}{|l|c|c||c|c|}
\hline \hline 
{} & \multicolumn{2}{c||}{\bf CLIC 3 TeV} & \multicolumn{2}{c|}{\bf CLIC 0.5 TeV} \\
\hline
Collimator & $a_x$ [mm] & $a_y$ [mm] & $a_x$ [mm] & $a_y$ [mm] \\
\hline \hline
ENGYSP (E spoiler) &  3.51 (2.5) & 8.0 & 3.0 & 12.0  \\
ENGYAB (E absorber) &  5.41 (4.0) & 8.0 & 4.6 & 12.0 \\
YSP$\#$ ($\beta_y$ spoiler) & 8.0 & 0.1 & 12.0 & 0.19 \\
YAB$\#$ ($\beta_y$ absorber) & 1.0 & 1.0 & 1.0 & 1.0 \\
XSP$\#$ ($\beta_x$ spoiler) & 0.12 & 8.0 & 0.39 & 12.0 \\
XAB$\#$ ($\beta_x$ absorber) & 1.0 & 1.0 & 1.0 & 1.0 \\
\hline
\end{tabular}
\end{center}
%\end{sidewaystable}
\end{table}

\noindent Concerning collimator wakefields, for both CLIC at 3 TeV CM and CLIC at 0.5 TeV CM, considering the beam parameters of Table~\ref{CLICparametros} and the collimator (spoiler) parameters of Table~\ref{tablecoll6}, the geometric wakefields (from Stupakov's criteria from Eq.~(\ref{kickgeom1})) are in the diffractive regime, near the border with the intermediate regime. 

Taking into account the dependence of the resistive wake kick on the beam parameters and the collimator aperture (see Eq.~(\ref{kickres})), $\langle y' \rangle \propto N_e/(E \sqrt{\sigma_z} a^3)$, the resistive kick from the vertical betatron spoilers for 0.5 TeV CM is approximately a factor $1.25$ larger than the kick for 3 TeV CM, $\langle y' \rangle_{0.5~{\rm TeV}}/\langle y' \rangle_{3~{\rm TeV}} \approx 1.25$. On the other hand, for the horizontal betatron spoilers the resistive kick ratio is $\langle x' \rangle_{0.5~{\rm TeV}}/\langle x' \rangle_{3~{\rm TeV}} \approx 0.25$.   

No simulations have yet been carried out for the collimation performance study of the CLIC optics at 500 GeV CM. In this regard further work is needed.        

\section{Summary, conclusions and outlook}

The post-linac collimation system of CLIC must fulfil two main functions: the minimisation of the detector background at the IP by the removal of the beam halo, and the protection of the BDS and the interaction region against miss-steered  or errant beams.   

Recently several aspects of the CLIC post-linac collimation system at 3 TeV CM energy have been optimised in order to improve its performance. This report has been devoted to explain the optimisation procedure and to describe the current status of the CLIC collimation system.   

The CLIC collimation system consists of two sections: one for momentum collimation and another one for betatron collimation. Next, the conclusions for the betatron collimation system are summarised: 

\begin{itemize}
\item The main function of the betatron collimation system is to provide the removal of those particles from the beam halo which can potentially contribute to generate experimental background at the IP.

\item Beam tracking simulations have shown optimum betatronic collimation depths at $15~\sigma_x$ and $55~\sigma_y$. For these depths the tracking simulations of a primary halo through the BDS have shown a good collimation efficiency of the system. 

\item An optimisation of the phase advance between the betatron spoilers and the final doublet has led to an additional $20\%$ improvement of the cleaning efficiency.

\item The betatron spoilers have to be set to relatively very narrow gaps ($\sim 100~\mu$m) for efficient scraping of the transverse beam halo. Therefore, the surface of the jaws of these spoilers are very close to the beam axis, and can significantly contribute to the luminosity degradation by wakefields when the beam pass through them with a certain offset from the nominal beam axis. The luminosity loss due to collimator wakefields has been computed, using the codes PLACET \cite{placetoctave} and GUINEA-PIG \cite{guinea}, and found to amount to up to $20\%$ for vertical beam offsets of $\approx 0.4~\sigma_y$. For this calculation spoilers made of Be have been assumed. This study has to be extended to other possible material options. 

\item Reducing the taper angle the geometrical contribution of the collimator wakefields is reduced. However, for the CLIC spoilers the resistive part of the wakefields is dominant, and only a very modest improvement in the minimisation of the wakefield effects has been found by reducing the taper angle to approximately 15 mrad. This translates into a longer spoiler (of almost 1 m) than the original 88~mrad spoiler (of 25 cm). Longer spoilers introduce tighter tolerances in terms of alignment and tilt errors. Therefore, we have finally decided to maintain the original taper angle of 88~mrad.

\item  For CLIC the betatron spoilers have always been assumed to be made of Be. The main arguments to select Be were its high thermal and mechanical robustness and good electrical conductivity (to minimise resistive wakefields). Nevertheless, an important inconvenience is the toxicity of Be-containing dusts, and accidents involving Be might be a serious hazard. Since no survivability to the full beam power is demanded for the betatron spoilers (they are designed to be sacrificial or consumable), the robustness requirement of the material could be relaxed and different options other than Be could be taking into account. For example, Ti-Cu coating or Ti alloy-Cu coating could be good candidates.       
\end{itemize}

\noindent For the collimation efficiency studies here we have assumed the spoilers as perfect collimators or `black' collimators, considering the particles of the primary beam halo perfectly absorbed if they hit a spoiler or a limiting aperture in the BDS. In this simplification no secondary production have been assumed. However, in order to make more realistic simulations, the performance of the optimised CLIC collimation system has to be studied using specific simulation codes for beam tracking in collimation lattices, such as BDSIM \cite{BDSIM1}. The tracking code BDSIM allows us to make a more realistic collimation scenario adding the production of secondary particles and its propagation along the lattice when a particle of the primary halo hit one spoiler or other component of the lattice. Recently an interface BDSIM-PLACET \cite{BDSIM2} has also been developed for the tracking of the beam halo through the BDS of linear colliders, including the wakefield effects and the production of secondaries. In addition, simulations using a more realistic model of the transverse halo would also be convenient. In this direction, notable progress has been made during the last years on the investigation and simulation of different mechanisms which generate transverse halo in both linac and BDS of the linear colliders. The code PLACET incorporates a module called HTGEN \cite{HTGEN}, which permits the simulation of the production of beam halo by beam-gas scattering and the tracking of this halo and the beam core along the lattice. For a more complete characterisation, we plan to apply all these simulation tools to the optimised collimation system.         

Measurements of collimator wakefields will be useful to validate the analytical and simulation results. In the past, sets of measurements have been made for longitudinally tapered collimators at SLAC End Station A (ESA), see for example \cite{PT3}. For the geometric wakefields, these measurements showed an agreement at the level of $20\%$ with the simulation results and good qualitative agreement with the theory, although in many cases there was a quantitative discrepancy as large as a factor 2 between theory and measurement. Measurements of the resistive wakefields \cite{PT4} showed notable discrepances with theory. New sets of measurements would be helpful, using available beam test facilities, such as ATF2 \cite{ATF2}, ESTB (former ESA) \cite{ESTB}, CALIFES \cite{Califes} and FACET \cite{FACET}. For instance, a possibility would be the use of the test facility FACET at SLAC, which will operate with longitudinally short bunches (20~$\mu$m bunch length) and bunch charge (1 nC) close to those of CLIC (44~$\mu$m bunch length and $0.6$ nC bunch charge).

For the CLIC option at 500 GeV CM energy the collimation system design is still in a premature state. In this sense, further work has to be made for its optimisation and consolidation.       
 
\section*{Acknowledgements}
This work is supported by the European Commission under the FP7 Research Infrastructures project EuCARD, grant agreement no. 227579.

\end{document}